\documentclass[preprint,aps]{revtex4}
\usepackage{graphicx}
\usepackage{booktabs}
\usepackage[usenames, dvipsnames]{color}
\usepackage{multirow}
\usepackage{eurosym}
\usepackage{setspace}
\usepackage{rotating}
\usepackage[ pdftex, plainpages = false, pdfpagelabels, 
                 pdfpagelayout = useoutlines,
                 bookmarks,
                 bookmarksopen = true,
                 bookmarksnumbered = true,
                 breaklinks = true,
                 linktocpage=all,
                 pagebackref=false,
                 colorlinks = true,
                 linkcolor = BrickRed,
                 urlcolor  = blue,
                 citecolor = BrickRed,
                 anchorcolor = green,
                 hyperindex = true,
                 hyperfigures
                 ]{hyperref}

\begin{document}

\title{\href{http://www.necsi.edu/}{Corporations and Regulators: \\ The Game of Influence in Regulatory Capture}} 
\date{\today}  
\author{Dominic K. Albino, Anzi Hu, and \href{http://necsi.edu/faculty/bar-yam.html}{Yaneer Bar-Yam}}
\affiliation{\href{http://www.necsi.edu}{New England Complex Systems Institute} \\ 
238 Main St. S319 Cambridge MA 02142, USA}

\begin{abstract}
In a market system, regulations are designed to prevent or rectify market failures that inhibit fair exchange, such as monopoly or transactions with hidden costs. Because regulations reduce profits to those possessing unfair advantage, corporations (whether individuals, companies, or other collective organizations) are motivated to influence regulators. Regulatory bodies created to protect the market are instead co-opted to advance the interests of the corporations they are charged to regulate. This wide-spread influence, known as ``regulatory capture,'' has been recognized for over 100 years, and according to expectations of rational behavior, will exist wherever it is in the mutual self-interest of corporations and regulators. Under the assumption of economic rationality, a theoretical analysis suffices to determine where regulatory capture will occur. Here we model the interaction between corporations and regulators using a new game theory framework explicitly accounting for players' mutual influence, and demonstrate the incentive for collusion. Communication between corporations and regulators enables them to collude and split the resulting profits.  We identify when collusion is profitable for both parties. The intuitive results show that capture occurs when the benefits to the corporation outweigh the costs to the regulator. Under these conditions, the corporation can compensate the regulator for costs incurred and, further, provide a profit to both parties. In the real world, benefits often far outweigh costs, providing large incentives to collude and making capture likely even when strict rationality may not apply. Regulatory capture is inhibited by decreasing the influence between parties through strict separation, independent market knowledge and research by regulators, regulatory and market transparency, regulatory accountability for market failures, widely distributed regulatory control, and anti-corruption enforcement. We discuss the impact of integrity of the regulator in the analysis, relaxing the rationality assumption. We also discuss various ways regulators themselves may seek opportunities to benefit.
\end{abstract}

\maketitle

The influence of corporations on government has been a concern of public leaders for over 100 years \cite{Wilson1961} and is a central concern of protest movements today \cite{SunlightFoundation2006, UnitedRepublic2011, Occupy2011}. When public interest is sacrificed to benefit a corporation (defined as an individual, company, or other collective organization) due to that corporation's influence on its regulators, the phenomenon is known as regulatory capture \cite{LaffontTirole1993, Bo2006, Boehm2007, CarpenterMoss2010}. The negative effects of regulatory capture have historically been acknowledged by both policy makers and economists \cite{Olson1965, Stigler1971, Tullock1980, LaffontTirole1993, SchleiferVishny1994, EstacheMartimort1999, Duso2005}, but the global financial crisis beginning in 2007 crystalized the idea that regulatory failure can threaten basic market function and the economy as a whole. The report of the Financial Crisis Inquiry Commission concluded that the crisis was avoidable and that ``widespread failures in financial regulation and supervision proved devastating to the stability of the nation's financial markets'' \cite{FCIC2011}. Given the essential role of regulation, it is critical to understand how capture causes regulatory failure, how widespread capture is, and how capture can be prevented or mitigated.

Here we analyze regulatory capture using a novel game theory framework with two players, the corporation and the regulator. Game theory is used to analyze the choices of players given payoffs that result from combinations of their actions. An essential aspect in our treatment is allowing influence between the players, rather than using the common assumption of independent players. Without communication, capture is not likely because the conditions, threats, or payoffs of collusive agreements cannot be conveyed. We analyze capture and identify its causes, conditions, and consequences. Economic rationality implies that when incentives for capture are large, it can be expected to occur. By identifying when incentives are large, we can identify with high certainty when capture affects regulatory decisions. Our results show the most important condition for regulatory capture is that the benefit of capture to the corporation is larger than the cost of capture to the regulator (typically composed of the risk of civil, criminal, and social penalties). When the benefit is much larger than the cost, the corporation can easily compensate the regulator more than enough to outweigh the cost, creating a profit for both the regulator and the corporation. In practice, corporate benefits can easily run into the billions of dollars \cite{Alexander2009, Drutman2012, Jilani2012}, whereas costs to the regulator rise higher than the millions of dollars only in exceptional circumstances. Because of this large discrepancy between the scale of benefits and the scale of costs, widespread capture can be expected. 

Among the conditions that reduce the likelihood of capture are: isolation of the regulators from the corporations (both at the time of regulatory oversight and over more extended time frames), increased visibility of the consequences of regulations, increased visibility of interactions between regulators and corporations, and widely distributed regulatory control. Conversely, corporations and regulators have strong incentives to increase interactions; make both market activities and regulatory actions more obscure and inaccessible; and centralize oversight, reducing the number of people involved. 

Framing the context for regulatory capture requires understanding why markets need regulation in the first place---which is sometimes contested by those claiming that free markets are essentially self-optimizing and self-regulating \cite{Hardy2006}. Underlying this claim are the premises that transactions are voluntary and that both immediate and long term consequences of every transaction are known to the participants. In a complex world, there are many causes for departures from these assumptions, including when transactions are not truly voluntary, when consequences and costs are not or cannot be known, and when there are important implications for anyone not immediately involved (externalities). 

Economic theory and practice have demonstrated the benefits of markets for coordinating economic transactions  \cite{Smith1776,Walras1877,Hicks1939,VonMises1941,ArrowDebreu1954,SamuelsonNordhaus2009-2}. However, undesirable outcomes such as monopoly, fraud, and pollution can still occur within a market framework \cite{Bator1958,GreenwaldStiglitz1986}. As a result, consumers face a variety of problems such as price-gouging, shoddy building construction, and the deleterious health effects of pollution. Economists label these consequences the results of market failure---such problems do not exist in ideal theoretical markets, but real markets do not behave in the idealized fashion commonly described in economic theory \cite{Stiglitz2009}. Market failures result from imperfect competition \cite{Marshall1920,Williamson1972,Lande2007}, information advantages \cite{Akerlof1970,Cseres2006,BrixMcKee2009} and information imperfections \cite{GreenwaldStiglitz1986}, externalities, transaction and information costs, and suboptimal equilibria \cite{Arrow1969}. 

Because of these problems, government or non-governmental organizations frequently attempt to improve market function through policy and regulation \cite{Medema2007}. To preserve the constructive role of markets, such policies usually avoid directly mandating economic transactions or prices, but instead attempt to restrict undesirable behavior, such as dumping barrels of toxic waste \cite{Stiglitz2009}. Policy-making bodies may improve market outcomes over the long run by establishing conditions closer to theoretical ideals, which include perfect information, competition, and rationality; complete and private property rights; and negligible transaction costs and externalities \cite{FullertonStavins1998}.  Some examples of regulation considered highly successful in U.S. history have been anti-trust regulation including the Sherman Antitrust, the Clayton Antitrust, and the Federal Trade Commission Acts \cite{Stigler1968}, enforced in part by the Department of Justice and the Federal Trade Commission, and the Clean Air and Clean Water Acts \cite{EPA2010,EPA1997,EPA2012}, enforced by the Environmental Protection Agency. 

Given the presence of regulation, corporations have strong incentives to influence what policies are implemented. Corporations
may benefit from the existence of market imperfections as well as the presence or absence of specific rules \cite{Stigler1971,LaffontTirole1992,CarpenterMoss2010}.
When regulating bodies are induced to favor specific market participants over the market as a whole, the purpose of regulation--to promote optimal
market function--is subverted, wholly or in part, resulting in regulatory capture.

Captured regulators fail to enact or enforce equitable policies---and may go so far as to actively protect the interests of their captors from other regulators---leaving the market under-regulated or misregulated. Extant but ineffective regulation may conceal the damage, making it difficult to address and resulting in conditions particularly harmful to society.

Opportunities for regulatory capture arise naturally from the way regulations develop.  When the U.S. government decides to regulate a market, it typically creates an agency vested with the necessary power and responsibility, such as the Securities and Exchange Commission, the Federal Communication Commission, the Food and Drug Administration, or the Environmental Protection Agency.  In order to support and protect well-functioning markets, regulators require extensive market information, and communication between regulators and corporations is often considered necessary for this purpose \cite{McCarty2011}. These ongoing interactions, however, also allow communication that reveals intended future actions and facilitates agreements on reciprocity, leading to the kind of influence  seen in collusion.

The key to characterizing regulatory capture is understanding the costs and benefits to corporations and regulators, as well as the interactions between the two parties. The benefit to the corporation from collusion is the difference between its financial outcome under favorable and unfavorable regulatory regimes.  The corporation's profit is its benefit less its cost, which is the amount the corporation must expend in order to influence the regulator, for example by direct transfer.  The regulator's profit is the amount paid to it by the corporation, less the costs the regulator faces for effecting biased regulation. In the simplest case, by totaling the incentives of the corporation and the regulator and assuming collusion will occur where profitable, a cost-benefit analysis indicates our primary conclusion that regulatory capture results from a high ratio of corporate benefits to regulator costs.

Regulatory capture typically requires decisions favorable to corporations from key individuals, including administrators and commissioners of regulatory agencies, and legislators and executives passing laws.  Financially motivated corporations may exert influence on regulation in several ways. Illegal, direct bribes or threats can cause individual regulators to compromise their opinions, as can legal, indirect favors for friends or relatives, donations to political campaigns and valued causes,  or richly compensated employment outside of regulated contexts. Decisions can also be influenced by biased information (lobbying) that reshapes the opinions of regulators or even those of the public; bringing to bear social pressure to adopt a favorable world view; or by the appointment of favorably biased individuals to powerful positions, arranged by influencing those able to make the appointments. 

Counter to the incentives provided by corporations attempting to influence regulation, regulators face moral considerations as well as the risk of exposure, with consequent criminal liability, loss of public career prospects, and damaged personal and professional reputation \cite{Woodward1998}. Most of the cost components are contingent on collusion being detected and exposed, so it is reasonable to assume costs will increase with the increased risk of detection associated with higher values and frequencies of transfers. It is also reasonable to assume a regulator who frequently acts in the corporation's interest is more likely to be detected and faces higher costs than one who only does so occasionally.  When detection risk is mitigated by using indirect methods to accomplish the transfer---payments or jobs for relatives or friends, contributions to a valued cause or political campaign---the benefit to the regulator must be discounted appropriately. Many of the regulator's costs are non-financial, such as social pressure or moral consideration, but we may be able to incorporate their effects with an appropriately adjusted financial proxy, simplifying the analysis to the single dimension of financial cost.

Corporations can be expected to exert the greatest effort where incentives are greatest---that is, where the most profit is at stake among the fewest parties \cite{Olson1965}. Regulators can be influenced through their desire for personal gain or their fear of harm to reputation or person \cite{Krueger1974}. Previous regulatory capture models \cite{Bo2006,Boehm2007} propose a wide range of factors that may be influential, including the advantage gained by successful captors \cite{Peltzman1976}; the impact of differential information across corporations, regulators, and the public \cite{Tirole1986,LaffontTirole1993}; capture strategies of monopolists versus those of competitive market participants \cite{GrossmanHelpman1993,Dixit1997,Tullock1980}; effects of multiple regulators on the behavior of both corporations \cite{Snyder1991} and regulators themselves \cite{Bo2000,Bennedsen2002,Neeman1999}; and the ``revolving doors'' phenomenon, in which many regulators work for corporations before or after their work as regulators \cite{Che1995, Overby2011, Vidal2011}.

We focus our attention on the importance of the interaction between regulators and corporations, and our resulting analysis of the incentives describes a game where cooperation arises from the mutual influence between the players---a ``collusive game". In order to make the importance of interaction clear, consider the case where a regulator and a corporation do not interact. Without interaction, the corporation assumes the regulator will behave strictly in its own best interest and take a proffered bribe without bothering to bias the regulation. Therefore, the corporation will not choose the useless cost of bribing. The regulator likewise assumes the corporation will act strictly in its own best interest and not reward favoritism with a kickback. Therefore the regulator will not choose the useless risk of biased regulation. This aligns with the mathematically equivalent example of the prisoners' dilemma, where individuals choose non-cooperation because it is a superior choice in the short run regardless of the other party's decision. Non-cooperative game theory provides an analysis in which each party makes a decision considering the best outcome for itself alone, resulting in the Nash equilibrium of the game where neither party chooses to cooperate.

Various complications of the prisoners' dilemma may alter the outcome.  In iterated prisoners' dilemma games, information about prior actions informs current actions and increases the benefits accruing to actors that practice selective cooperation \cite{Axelrod1981, Nowak1992, Nowak1993, Nowak1995}. Other work supports the idea that communication, both direct and indirect, can have an important impact on outcomes through reputational effects and reciprocity \cite{Gouldner1960, Nowak1998, Berg1995}. The importance of reciprocity has been observed in empirical studies \cite{Berg1995, Fehr2000, Clark2001, Lambsdorff2010}. Important assessments of communication in game theory include the analysis of Mutually Assured Destruction \cite{Schelling1960} in which various communications of one party influence the decisions of the other while engaged in nuclear confrontation. In the real world, direct communication, indirect communication, and continuing relationships between regulators and corporations are the norm. Rather than employ the specific frameworks of iteration, reputation, or reciprocation, we consider a mathematical framework that directly describes the influence between players.

Our treatment begins from the recognition that interactions between regulators and corporations facilitate exchange of information about market function and provide opportunities for communication that violate the typical independence assumptions of non-cooperative game theory. In order to incorporate their effect, we characterize the extent to which one party's decisions affect those of the other party: when one party acts to benefit the other, to what extent can it expect cooperation? We call this the degree of influence. When the parties estimate that the degree of influence times the benefit of collusion is greater than the cost of collusion, they expect gains from cooperation.

Our model is a two-player game structured as follows: in a certain regulatory process, the corporation can gain a benefit, $B$, from one of two regulatory outcomes. The regulator faces a cost, $C$, to choose the outcome that benefits the corporation. To influence the regulator, the corporation can offer a transfer, $t$, which becomes a cost for the corporation and a benefit for the regulator. The regulator allows the corporation a degree of influence, $\Delta$, which is the amount of increase in the probability of a favorably biased decision. The public's influence on the regulator is modeled implicitly by the regulator's expected cost from collusion. The central result of the analysis is that when the degree of influence times the corporation's benefit is larger than the transfer, which is in turn larger than the degree of influence times the regulator's cost, collusion makes both the corporation and the regulator better off and is the optimal choice for both players in an ongoing relationship (see the Appendix for technical details),

\begin{equation}
\Delta B>t>\Delta C.\label{eq:J_11cond}.
\label{eq:cond}
\end{equation}

When the effective cost to the corporation of the regulatory influence, $t_c$, is higher than the benefits received by the regulator, $t_r$, due to indirect costs and discounting of indirect transfers, the inequalities separate to become $\Delta B>t_c$ and $t_r>\Delta C$.

\begin{figure}
\includegraphics[scale=0.80]{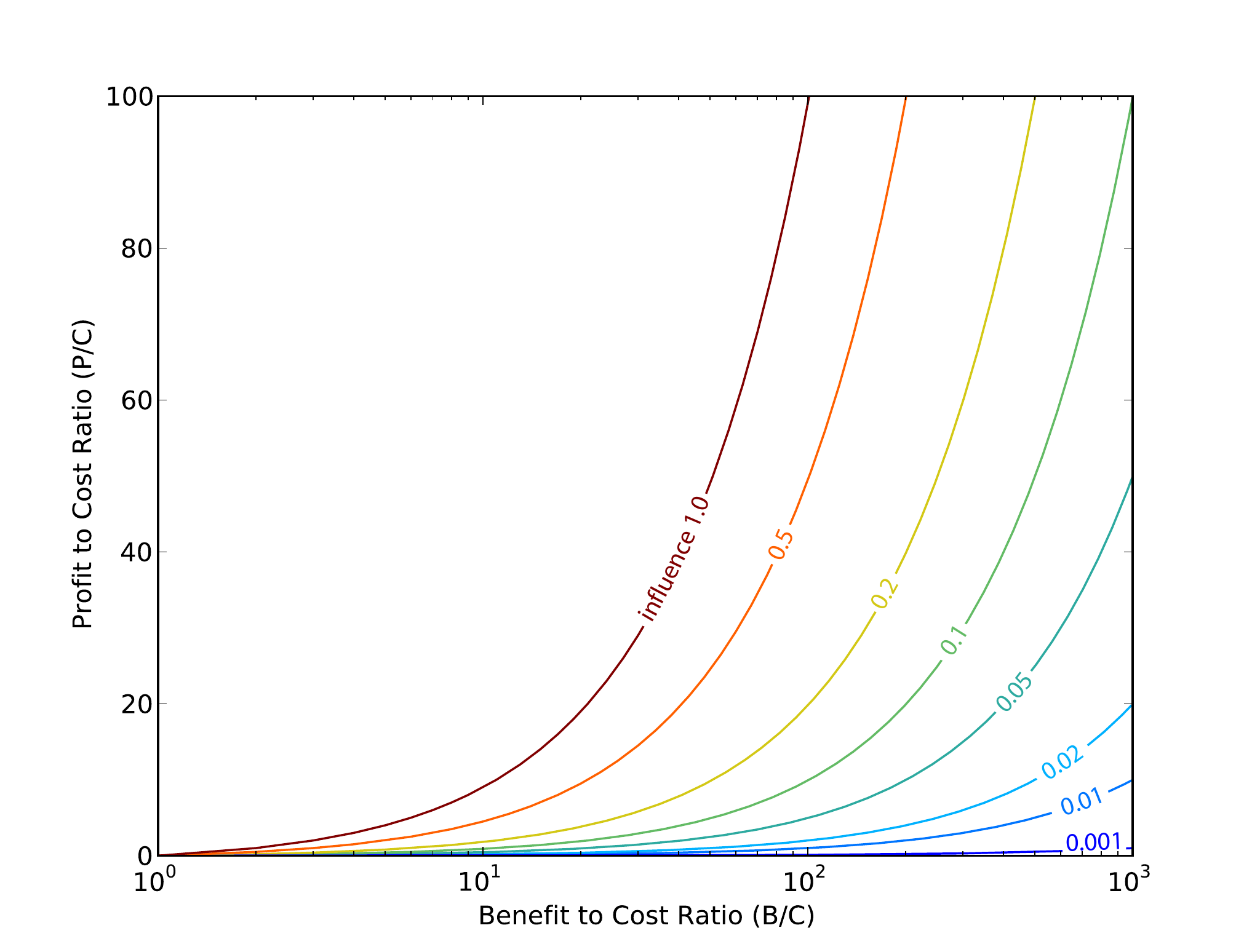}
\caption{A plot of colluders' realized profit as a function of their potential benefit over varying degrees of influence from 0.001 to 1.0. We normalize by colluders' cost. This relationship is linear, but we show it on a linear-log plot to emphasize that colluders' benefits can accrue at industrial scales while their costs are bounded by human scales.  Because of this difference, benefits grow dramatically larger than costs, driving large collusive profits and creating extremely strong incentives for capture.
}
\label{profit}
\end{figure}

More generally, Figure (\ref{profit}) shows that as the difference rises between the corporation's benefit and the regulator's cost, and as the degree of influence rises, so does the profit available to colluders, and therefore the incentive for regulatory capture. The regulator and corporation must negotiate the allocation of the resulting profits. To induce the regulator to collude, the corporation must offer a transfer large enough to more than offset the regulator's cost. Beyond the regulator's break-even point, the negotiation divides the profit between the parties, whose interests are opposed. Although the exact profits for each party depend on the split, total profit is a good measure of the combined incentives for collusion because it is an upper bound on the profit available to either player. 

Regulatory capture may be partial rather than total, allowing influence over some but not other decisions or swaying a decision partially but not fully in the corporation's favor \cite{Carpenter2011}. By considering the influence parameter, the model can also be used to assess partial influence over regulatory decisions. Under conditions of partial influence, the expected value of costs and therefore the minimum size for transfers are reduced, leading to possible collusion with smaller benefits and transfers.

We may estimate the regulator's cost, $C$, as 

\begin{equation}
C = \Delta (RA + I)\label{eq:CostFn}
\label{eq:cost}
\end{equation}

in terms of the amount of influence granted, $\Delta$, the risk of detection, $R$, the regulator's personal assets at risk, both financial and social, $A$, and the personal integrity of the regulator, $I$.  This expression indicates cost increases with the amount of influence being granted, the level of risk associated with collusion, the assets the regulator stands to lose, and the regulator's integrity.  When the regulator acts in a strictly rational way, integrity is zero and decisions are based strictly on the risk-reward assessment of economic benefits.  As the regulator's level of integrity increases, the analysis departs from the strictly rational limit and into a behavioral paradigm. At the level of approximation used in this expression, a behavioral model can be converted to a rational model by adjusting the regulator's cost---increasing integrity increases the effective cost.  For high levels of integrity, these costs can become large enough to outweigh the benefits of collusion and impede regulatory capture.  Levels of integrity vary across individuals and cultures, and may reflect, for instance, the structure of social loyalties and influences, the effects of public shaming \cite{Kahan1999,Barr2001,Schulze2003,Panagopoulos2010,Krain2012}, or the magnitude of external threats \cite{Bar-Yam2005}.  For example, the Cold War era may be expected to have given rise to higher levels of integrity in the US than the current one, where there is no comparable external threat. 

Risk, $R$, plays a key role in the regulator's cost, and a major component of the risk is the transparency of regulation and its effects.  Since the costs of collusion increase with the risk of detection, as it becomes clearer that a decision favors a corporation over the public interest, the risk and thus cost rise for colluders.  Advances in scientific knowledge and literacy can improve regulatory outcomes by raising public knowledge of the effects of regulation, making collusion easier to identify, increasing the risk of detection and thereby the costs of collusive regulation. 

The effects of transparency and moral principle can be illustrated by the distinction between regulation of public health and regulation of finance.  In a case of bacterial outbreak in a grocery product, many deaths could result if the regulator conceals the problem and protects the producer. During an outbreak of food-borne illness, there is high profile media coverage and transparent effects. Such public exposure makes the risk of detection and thus the difficulty of collusion high. The regulator faces the high moral cost of clear responsibility for innocent deaths. Even though benefits to the corporation might be large, they may be lower than the cost to the regulator. The regulator is thus unlikely to favor the corporation and will act instead in the best interest of the public. In contrast, in a case of regulation of financial markets such as complex derivatives, the details of the case are often unintelligible to the public. Even if regulations could be made transparent, they are often designed to obscure instead. As we have seen from the financial crisis in 2007-2008 and speculation in food commodity markets \cite{Lagi2011a,Lagi2011b}, life and health repercussions from market imperfections exist, but are obscured by cascading effects and time delays and thus poorly recognized.  Economic consequences may be inappropriately dismissed based on claims about the self-regulatory function of markets \cite{Hardy2006}. This drastically reduces the costs to colluders from both public pressure and perceived moral obligation. 

Specific assumptions about the composition of the regulator's cost do not affect the function of our model, which takes the regulator's costs in a particular circumstance as given (exogenous). In the case of an agency or other body like Congress composed of multiple individuals, the costs for the group are an appropriately defined aggregate of individual costs, keeping in mind that voting bodies can be influenced through their least costly members. 

In the real world, the benefits to corporations are often on industrial scales characteristic of markets large enough to justify regulation, but costs to the regulator are on human scales characteristic of individual salaries---as found, for example, in highly paid positions offered by corporations to regulators after their regulatory tenure. Moreover, as the strength of the financial incentive increases, the relative strength of non-financial considerations, including ethics, declines. With benefits that are often greater than costs by orders of magnitude, the incentives for regulatory capture are overwhelming. 

A 2009 case study \cite{Alexander2009} documented the profits and expenditures of firms lobbying for the American Jobs Creation Act of 2004, which granted certain corporations the opportunity for a large one-time tax benefit. The authors conservatively estimate firms spent a total of \$282.7 million lobbying for this bill and received \$62.5 billion in tax savings from it, a return on investment in excess of 220:1. 
The regulators' cost is less than the transfer from the corporations (i.e. the lobbying expenditure), and thus the ``purchase price" of favorable legislation is several orders of magnitude smaller than the benefits obtained. The bill passed the House of Representatives with a vote of 280-141 \cite{House2004} and the Senate with a vote of 69-17 \cite{Senate2004}. This amounts to more than \$530,000 spent per member of Congress (433 Representatives and 100 Senators) or \$810,000 spent per favorable vote. The Speaker of the House, the highest paid member of Congress, earned an annual salary of \$223,500 and most representatives and senators earned salaries of \$174,000 \cite{housepay,senatepay}.  While members of Congress did not receive the full amounts of the lobbying, these numbers demonstrate an empirical upper bound on the amount needed to achieve substantial influence, which is on the order of the regulators' salaries.

Our quantitative model provides additional insights into real world regulation and regulatory capture. We note that regulators and corporations cooperate in order to generate collusive profits, but compete over the potentially uneven share each ultimately obtains. Various cultures or contexts may determine differently how the benefits are allocated. Further, the analysis shows a market for regulatory capture can arise in which regulators compete to be captured in order to gain the associated benefits. Similarly, corporations can compete to achieve influence and be the ones to benefit from a captured regulator. Moreover, while it is commonly assumed that corporations are the driving force behind collusive relationships with regulators and regulatory capture, there is similarly motivation for regulators to seek such relationships. Regulators can incentivize corporations to provide benefits by enacting regulations or pursuing spurious enforcement actions that are detrimental to the corporation. Where there is no transparency, such actions can result in effective coercion for companies to ``play the game'' even where they would not otherwise do so.

Agencies with poor transparency, revolving doors, few commissioners, and limited accountability invite regulatory capture, and those standing to benefit from capture have strong incentive to encourage such conditions. On the other hand, regulatory bodies that avoid these pitfalls are more likely to achieve effective regulatory outcomes, especially by creating life-long career prospects and an independent, research-oriented approach. Most anti-corruption measures attempt to monitor direct transfers from corporations to regulators, which are easily circumvented. Such efforts increase the costs to the corporation, $t_c$, relative to the benefits received by the regulators, $t_r$, so the corporation pays more than the regulator receives, $t_c>t_r$, but this depreciation is typically small compared to the overall benefit of collusion and is unlikely to alter the outcome. For this reason, our analysis shows direct transfers are an ineffective metric compared to the potential of monitoring corporate benefitss gained through favorable regulatory change \cite{Rose-Ackerman1978, Rose-Ackerman1999, Shleifer1993, Glaesera2006}. 

In many contexts, the recommended approach to protecting public interest is to isolate regulatory agencies from ``political'' influence. This assumes elected officials, without expert knowledge, will adopt harmful ideological objectives \cite{Epstein1999, Bo2006, Boehm2007, Carpenter2011}. Our analysis suggests that isolation from politics is not effective in protecting the public interest, because corporate capture of regulators is widespread and apolitical. Indeed, regulators who are less answerable to the public may more readily adopt the views of corporate interests because they have an indirect and tenuous connection the public and its interests. Political isolation fails to address the primary drivers of regulatory capture: the interaction between regulators and regulated industries and the incentive of large collusive profits.

Our analysis suggests regulatory capture is widespread, but the features identified above indicate certain steps that can mitigate capture and protect the independence of regulatory decisions. One is isolating regulators from the industry they control by enhancing the potential for lifelong regulatory careers and improving the information and expertise independently possessed by regulators, so as to decrease the need for communication with the industry. Another is increasing the transparency of regulation by promoting unbiased research to analyze and clearly report the consequences of any regulation or deregulation.  Other approaches are helpful, but unlikely to have a large impact by themselves. Such approaches include increasing the potential cost of capture to corporations by distributing control among more regulators and increasing the risk colluders bear by raising both penalties and the rate of enforcement.  Taken together, changes in regulatory practice may meaningfully alter the payout structure of the collusive game played by corporations and regulators, increasing cost, decreasing profit, and reducing incentive to collude.

In summary, our game theoretic analysis shows that when the expected benefit to the corporation is greater than the expected cost to the regulator, the corporation can transfer enough to the regulator to overcome the cost and create a profit for both parties, incentivizing such action. Under these conditions capture is likely. While the benefits to corporations may be publicly visible, the wide range of ways corporations can influence regulators are not easily exposed. The expected profit also depends on the degree of influence, which may result in partially modified statutes or favorable outcomes in some but not all decisions, under the guise of a ``compromise'' or ``moderate'' regulatory position.  Increased likelihood of punishment raises the cost to the regulator, but this increase is often insufficient because corporate benefits are on a vastly larger scale.  Further, regulators are rarely held accountable for inadequate or inappropriate regulation. Even when clear regulatory failures have come to light, consequences for the regulators responsible have been negligible \cite{Puzzanghera2008, Lawson2011, Taibbi2011a, Taibbi2011b, Ferguson2012}.

Where large differences exist between the benefits and costs of capture, increasing the costs to regulators through various methods, including increased enforcement, is not likely to have a significant impact; widening the distribution of regulatory control increases effective costs, but as the example of the American Jobs Creation Act shows, even hundreds of regulators do not change the balance of incentives for large markets. In some cases, regulators can be effectively isolated from interactions with corporations, but this requires vigilance before and after, not merely during, the regulatory process. For effective regulation, it is crucial to increase public visibility of the regulatory benefits accruing to corporations and use these benefits to identify regulatory capture. Since the corporate benefits are much larger than potential transfer payments to regulators, this signature of regulatory capture should be more easily observed.

\vspace{10 mm}
We thank Richard Cooper, Jeffrey Fuhrer, Casey Friedman, Yavni Bar-Yam, Julius Adebayo, Karla Bertrand and Ramon Xulvi-Brunet for helpful comments on the manuscript.

\newpage{}


\section*{Appendix}

We demonstrate the formation of collusion between the regulator and corporation using a two-player model.
We allow the regulator to choose a regulatory scheme that either favors
the corporation (F) at the cost of public interest, or does not favor
the corporation and preserves the public interest (NF). The corporation
may either attempt to influence the regulator (I) or not (NI). The
cost to the regulator for favoring the corporation is $C$, the benefit the corporation gains
from a favorable regulatory scheme is $B$, and the size of the transfer
the corporation makes in its bid for influence is $t$. The payoff matrix is shown in Table \ref{tab:payoff}.

\begin{table}[htpb]

\begin{tabular}{|c|c|cc|cc|}
\hline 
 & \multicolumn{1}{c}{} &  & \multicolumn{1}{c}{Corporation} &  & \tabularnewline
\hline 
 &  & Not Influence & (NI) & Influence  & (I)\tabularnewline
\cline{2-6} 
 & Not Favor &  & 0 &  & $-t$\tabularnewline
Regulator & (NF) & $0$ &  & $t$ & \tabularnewline
\cline{2-6} 
 & Favor &  & $B$ &  & $B-t$\tabularnewline
 & (F) & $-C$ &  & $t-C$ & \tabularnewline
\hline
 
\end{tabular}
\caption{Payoff table of profits for the corporation (upper right of each cell) and the regulator (lower left of each cell) for each of four conditions resulting from combinations of actions by the regulator to favor or not to favor the corporation, and the corporation to try to influence or not to try to influence the regulator. $B$ is the benefit to the corporation from regulatory favor, $C$ is the cost to the regulator of granting favor, and $t$ represents the value of the transfer from the corporation to the regulator in the attempt to influence regulatory action.}
\label{tab:payoff}

\end{table}

In a standard, non-cooperative game theory analysis, players are assumed to be independent, and
each player maximizes its own payoff given all possible actions of the other. 
Under such assumptions (and provided $B,C,t>0$) our model is equivalent to the Prisoners' Dilemma. 
If the corporation does not attempt to influence (NI) the regulator, not favoring the corporation (NF) is naturally preferable for the regulator,
because it avoids the cost, $-C$, of favoring (F) the corporation. If the corporation does attempt to influence (I) the regulator, not favoring
the corporation (NF) remains preferable for the regulator, since it can obtain the advantage, $+t$, of the corporation's decision without  having to
pay the cost, $-C$, in return. This means not favoring the corporation (NF) is always preferable for the regulator, regardless of the corporation's action. 

Similarly, if the regulator decides not to favor (NF) the corporation, the corporation stands to gain nothing by attempting to influence (I) the regulator except to incur a loss, $-t$, so the corporation prefers not to influence (NI).  If the regulator favors the corporation (F), the corporation gains the benefit, $+B$, regardless of its own action,  
and attempting to influence (I) would only impose a cost, $-t$, so the corporation still prefers not to attempt to 
influence (NI). Thus, not attempting to influence (NI) is always the better strategy for the corporation,
regardless of the regulator's action. Therefore, the Nash equilibrium
of this game is \{NF, NI\}: the regulator finds it is never beneficial but always costly to favor the corporation (and so never does) 
and the corporation finds it is never beneficial but always costly to attempt to influence the regulator (and so never does).

In the Prisoners' Dilemma, as in our case, the Nash equilibrium is not
the optimal payoff condition for the players, either individually or collectively. 
The expected Nash equilibrium outcome is not robust to modifications of the game such as
iteration \cite{Axelrod1981}, which can instead result in the collective (global) optimum.
In our case, if the benefit to the corporation is larger than the
transfer and the transfer is in turn larger than the cost to the regulator, $B>t>C$,
the global optimum is collusion, where the regulator favors the corporation and the corporation 
influences the regulator \{F,I\}.

In real world market conditions, cooperation is possible, and actions are not independent.
We assume communication and reputation effects exist which signal players' willingness
to cooperate, and therefore the action of one player can be conditioned on the action of the other. 
We formalize this cooperation by considering the regulator's action to be contingent on
the corporation's action according to a conditional probability matrix $P_{x|y}$,
where $x$ is the regulator's action (F or NF), and $y$ is the corporation's action (I or NI).
The conditional probability $P_{x|y}$ is the probability of the regulator choosing 
action $x$ given that the corporation chooses action $y$. Specifically,
$P_{F|I}$ is the probability of the regulator favoring the corporation
given that the corporation attempts to influence the regulator, and $P_{F|NI}$
is the probability of the regulator favoring the corporation when the
corporation does not attempt to influence the regulator. If the regulator
acts independently, the two probabilities are the same, indicating no influence. But if the
attempt to influence has some positive effect, $P_{F|I}$ is greater than $P_{F|NI}$.
We call the difference $\Delta=P_{F|I}-P_{F|NI}$ the corporation's degree of
influence on the regulator. Note that negative values are theoretically possible for $\Delta$,
indicating the attempt to influence causes an adverse reaction (e.g. moral outrage). 
Since the conditional probability satisfies the relationships

\begin{equation}
\begin{array}{ll}
P_{NF|I}&=1-P_{F|I}\\
P_{NF|NI}&=1-P_{F|NI}
\label{eq:J01}
\end{array}
\end{equation}
$P_{NF|I}$ and $P_{NF|NI}$ need not be additionally specified.

The expected utilities for regulator and corporation are written in terms of the probability
the corporation attempts to influence, $p$

\begin{equation}
\begin{array}{ll}
U_{R}& = -CP_{F|I}p-CP_{F|NI}(1-p)+tp \\
 & = -C(\Delta-\frac {t}{C})p-CP_{F|NI}
 \label{eq:U1}
\end{array}
\end{equation}

and 

\begin{equation}
\begin{array}{ll}
U_{C} &=BP_{F|I}p+BP_{F|NI}(1-p)-tp  \\
 & =B(\Delta-\frac {t}{B})p+BP_{F|NI}.
\label{eq:U2}
\end{array}
\end{equation}

Depending on the value of $\Delta$ relative to $\frac {t}{C}$ and $\frac {t}{B}$, the maximum value of the utilities
are at either $p=0$, i.e. not attempting to influence is strictly preferred, or $p=1$, i.e. attempting to 
influence is strictly preferred.
If $\Delta > \frac {t}{C}$, the regulator's utility function decreases with $p$, and for $\Delta < \frac {t}{C}$, it increases 
with $p$. In the latter case, effecting a transfer is favorable to the regulator. 
If $\Delta<\frac {t}{B}$, the corporation's utility function decreases with $p$, and for $\Delta>\frac {t}{B}$, it increases with
$p$. Thus, the transfer is beneficial to both the regulator and the corporation if 

\begin{equation}
\frac{t}{B}<\Delta<\frac{t}{C}.
\label{eq:Deltacondition}
\end{equation}
For positive $B, C, t$, this is equivalent to the inequality
\begin{equation}
\Delta C<t<\Delta B.
\label{eq:tcondition}
\end{equation}
Both sets of inequalities represent equivalent conditions for collusion. Inequality \ref{eq:Deltacondition} shows the range for the degree of influence that will generate cooperation given a fixed benefit, cost, and transfer. Inequality \ref{eq:tcondition} shows the range of transfer sizes that will generate cooperation given fixed benefit, cost, and degree of influence. The threshold condition for collusion is only satisfied when $t<\Delta B$, reflecting the requirement that a corporation would not provide a transfer larger than its realized benefit. Similarly, $\Delta C<t$ reflects the regulator's requirement that the payoff must be larger than the realized cost.  Under these conditions, both colluders will make a profit. 

The sum of the corporation's profit and the regulator's profit is equal to the amount of influence times the difference between the corporation's benefit at full influence and the regulator's cost at full influence, $\Delta(B-C)$. 
The colluders split this profit by bargaining over the size of the transfer, $t$, and the degree of influence, $\Delta$.  The regulator captures the proportion of the profit given by $\frac{t-\Delta C}{\Delta (B-C)}$ and the corporation captures the proportion given by $\frac{\Delta B -t}{\Delta (B-C)}$. 

The corporation benefits from collusion when the degree of influence, $\Delta$, is larger than the ratio between the transfer and the benefits, $\frac{t}{B}$, and in general wants the degree of influence to be as large as possible compared to this ratio. The regulator benefits from collusion when the transfer, $t$, more than compensates for the risk incurred by favoring the corporation, $\Delta C$, and prefers the difference between the transfer and the realized cost be as large as possible. Examining these incentives shows that during profitable collusion, players still have opposed interests in profit splitting. The split of profits is determined by both transfer size, $t$, and degree of influence, $\Delta$, with the regulator benefitting from a large transfer and small degree of influence while the corporation benefits from a small transfer and large degree of influence.

When either the transfer size, $t$, or the degree of influence, $\Delta$, is fixed, the other value may be used to apportion the agreed split of profits. Bargaining over the transfer, $t$, affects only the split of the profits, but bargaining over the degree of influence, $\Delta$, also affects the total profit available to split.  This is because partial degrees of influence, $0<\Delta <1$, lead to colluders collectively capturing only part of the potential profit ($\Delta B- \Delta C$ is maximized at $\Delta = 1$ for $B>C>0$).

\begin{figure}
\includegraphics[scale=0.8]{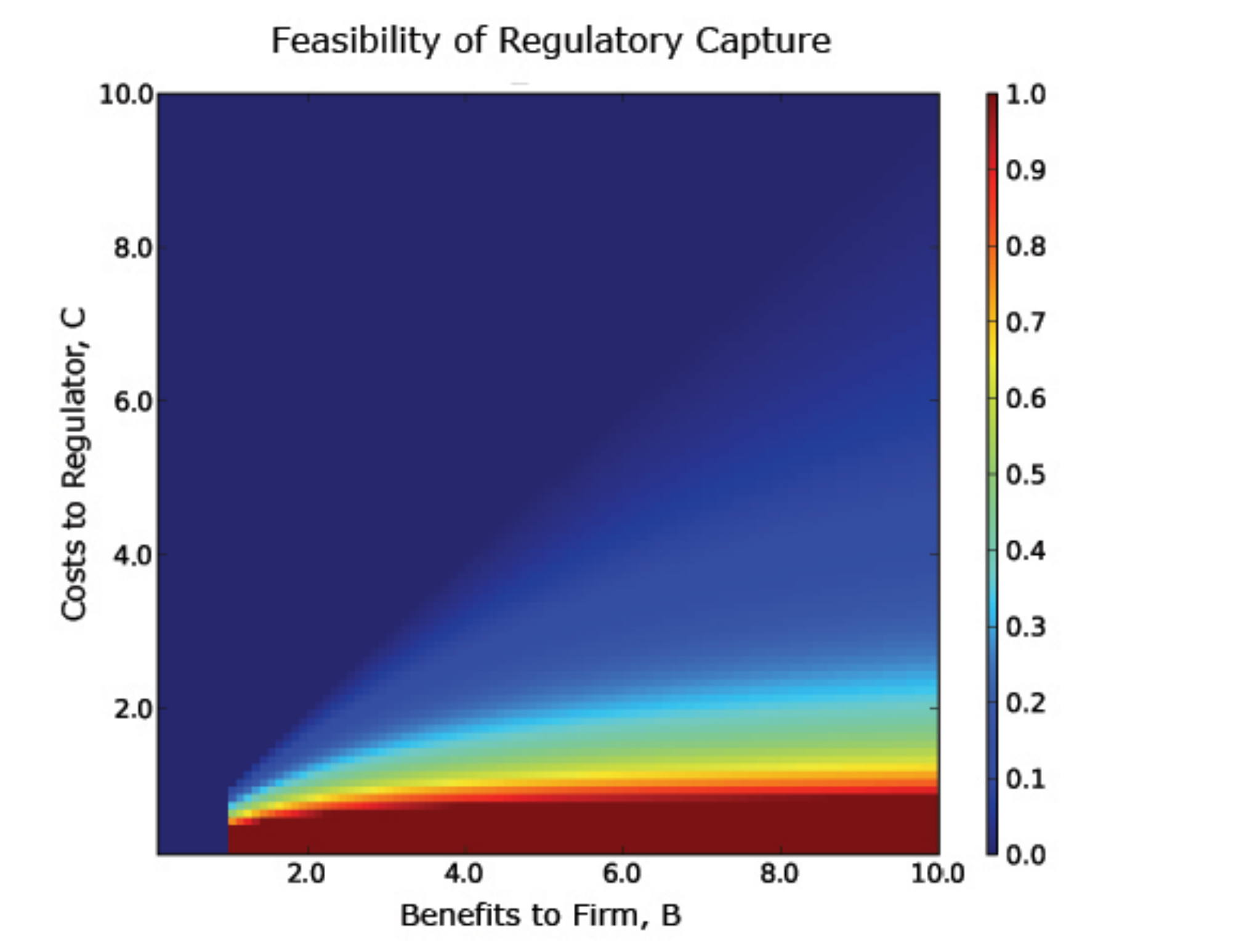}
\caption{The feasibility of regulatory capture, measured by the difference between the maximum and minimum degrees of influence leading to capture, $min{(\frac{t}{C},1) - \frac{t}{B}}$, given a constant size transfer of $t=1$.  The larger the difference, the larger the colluders' profit, and the stronger the incentive for regulatory capture.}
\label{fig:deltarange}
\end{figure}

In Fig. \ref{fig:deltarange}, we plot the size of the feasible range of $\Delta$, which is the difference between maximum and minimum degrees of influence leading to capture, as a function of the benefit and cost, $\frac{t}{B}-\min\{\frac{t}{C},1\}$. We assume a fixed transfer value without loss of generality. This represents a normalized measure of the aggregate profit accruing to colluders and thus their aggregate incentives for collusion.  In the red region the range is large and almost any positive amount of influence, $\Delta$, supports collusion. Therefore collusion is almost certainly taking place when the benefit, $B$, and cost, $C$, fall in that region. In this region the cost to the regulator is always offset by the transfer, $C<t$, and the regulator benefits from the collusion even under full influence, $\Delta=1$, so the regulator is always motivated to collude. For corporations, the transfer is much smaller than the benefits gained as $\frac{t}{B}$ approaches zero. In this region, a very small increased chance of favorable regulation---i.e. a very small degree of influence, $\Delta$---is sufficient to motivate the corporation to collude. 

Outside the red domain, the degree of influence, $\Delta$, has to satisfy more restrictive conditions for collusion to occur. An interesting case occurs when the transfer is smaller than the regulator's cost, $t<C$. By only partially or occasionally acting in favor of the corporations, the regulator can reduce the actual cost realized to less than the amount of the transfer, $\Delta C< t$. If a regulator is making a series of distinct decisions, all of which affect the outcome for the corporation, the regulator may favor the corporation with some decisions and not with others. Under such partial influence, the risks and resulting costs to the regulator and benefits actually achieved by the corporation are reduced, but may allow the players to reach mutually profitable collusion. 

Finally, it is also possible to describe collusion as an influence market in which the regulator dispenses (sells) influence, and the corporation seeks to buy it. The effective price of influence is $p=\frac{t}{\Delta}$. For collusion to be rational, this price has to be bounded by the benefits and the costs, 

\begin{equation}
C<\frac{t}{\Delta}<B.\label{eq:fraccondition}
\end{equation}
Thus, if the amount of transfer required to achieve influence is either larger than the benefit the corporation stands to gain, $B$, or smaller than the cost the regulator must face, $C$, the sale of influence will not occur. Using this interpretation we can consider a market in which different regulators are competing. A regulator would strive to increase the degree of influence, $\Delta$, in order to make a sale. This is characteristic of regulators signaling that they are easy to influence and attracting corporations. Similarly, corporations may compete to be in a better position to influence regulators and thus have an increased opportunity to buy. 

\newpage

\bibliographystyle{Science}
\bibliography{regcaprefs}

\begin{thebibliography}{10}

\bibitem{Wilson1961}
W.~Wilson, {\it The new freedom: A call for the emancipation of the generous
  energies of a people\/} (Prentice-Hall, Englewood Cliffs, New Jersey, 1961).

\bibitem{SunlightFoundation2006}
Sunlight Foundation, {\it Our mission\/}  (2006).
  \url{http://sunlightfoundation.com/about/}.

\bibitem{UnitedRepublic2011}
United Republic, {\it Our mission\/}  (2011).
  \url{http://unitedrepublic.org/about/our-mission}.

\bibitem{Occupy2011}
Occupy Wall Street New York City General Assembly, {\it Declaration of the
  occupation of New York City\/}  (2011).
  \url{http://www.nycga.net/resources/documents/declaration/}.

\bibitem{LaffontTirole1993}
J.~J. Laffont, J.~Tirole, {\it A theory of incentives in procurement and
  regulation\/} (The MIT Press, Cambridge, Massachusetts, 1993).

\bibitem{Bo2006}
E.~D. B\'o, Regulatory capture: A review, {\it Oxford Review of Economic
  Policy\/} {\bf 22}, 203 (2006).

\bibitem{Boehm2007}
F.~Boehm, Regulatory capture revisited -- Lessons from economics of corruption,
  {\it Working Paper\/}  (CIEP, Universidad Externado de Colombia, 2007).
  \url{http://www.icgg.org/downloads/Boehm%20-%20Regulatory%20Capture%20Revisited.pdf}.

\bibitem{CarpenterMoss2010}
D.~Carpenter, D.~Moss, Introduction, in D. Carpenter, D. Moss (eds.), {\it
  Preventing capture: Special interest influence in regulation, and how to
  limit it\/}  (Cambridge University Press, forthcoming 2013).

\bibitem{Olson1965}
M.~Olson, {\it The logic of collective action: Public goods and the theory of
  groups\/} (Harvard University Press, 1965).

\bibitem{Stigler1971}
G.~J. Stigler, The theory of economic regulation, {\it Bell Journal of
  Economics and Management Science\/} {\bf 2}, 3 (1971).

\bibitem{Tullock1980}
G.~Tullock, Efficient rent seeking, in J. M. Buchanan, R. D. Tollison, and G.
  Tullock (eds.), {\it Towards a theory of the rent-seeking society\/}  (Texas
  A\&M University Press, Texas, 1980).

\bibitem{SchleiferVishny1994}
A.~Shleifer, R.~W. Vishny, Politicians and firms, {\it Quarterly Journal of
  Economics\/} {\bf 109}, 995 (1994).

\bibitem{EstacheMartimort1999}
A.~Estache, D.~Martimort, Politics, transaction costs, and the design of
  regulatory institutions, {\it World Bank Policy Research Working Paper No.
  2073\/}  (1999).
  \url{http://elibrary.worldbank.org/content/workingpaper/10.1596/1813-9450-2073}.

\bibitem{Duso2005}
T.~Duso, Lobbying and regulation in a political economy: Evidence from the U.S.
  cellular industry, {\it Public Choice\/} {\bf 122}, 251 (2005).

\bibitem{FCIC2011}
Financial Crisis Inquiry Commission, The financial crisis inquiry report: Final
  report of the national commission on the causes of the financial and economic
  crisis in the {U}nited {S}tates  (Government Printing Office, Washington,
  D.C., 2011). \url{http://www.gpo.gov/fdsys/pkg/GPO-FCIC/pdf/GPO-FCIC.pdf}.

\bibitem{Alexander2009}
R.~Alexander, S.~Mazza, S.~Scholz, Measuring rates of return for lobbying
  expenditures: An empirical analysis under the American Jobs Creation Act,
  {\it Journal of Law and Politics\/} {\bf 25} (2009).
  \url{http://papers.ssrn.com/sol3/papers.cfm?abstract_id=1375082}.

\bibitem{Drutman2012}
L.~Drutman, Lobby more, pay less in taxes, {\it Sunlight Foundation\/}  (2012).
  \url{http://sunlightfoundation.com/blog/2012/04/16/lobby-more-pay-less-in-taxes/}.

\bibitem{Jilani2012}
Z.~Jilani, The amazing ROI of corporate lobbying, {\it United Republic\/}
  (2012).
  \url{http://www.republicreport.org/wp-content/uploads/2012/04/ROI.jpg}.

\bibitem{Hardy2006}
D.~C.~L. Hardy, Regulatory capture in banking, {\it International Monetary Fund
  Working Paper WP/06/34\/}  (2006).
  \url{http://www.imf.org/external/pubs/ft/wp/2006/wp0634.pdf}.

\bibitem{Smith1776}
A.~Smith, {\it An inquiry into the nature and causes of the wealth of
  nations\/} (W. Strahan and T. Cadell, London, 1776).

\bibitem{Walras1877}
L.~Walras, {\it Elements of pure economics\/} (American Economic Association,
  New York, 1877).

\bibitem{Hicks1939}
J.~Hicks, {\it Value and capital: An inquiry into some fundamental principles
  of economic theory\/} (Clarendon Press, Oxford, 1939).

\bibitem{VonMises1941}
L.~Von~Mises, {\it Interventionism: An economic analysis\/} (The Foundation for
  Economic Education, Irvington-on-Hudson, New York, 1998).
  \url{http://citeseerx.ist.psu.edu/viewdoc/download?doi=10.1.1.175.1861&rep=rep1&type=pdf}.

\bibitem{ArrowDebreu1954}
K.~J. Arrow, G.~Debreu, The existence of an equilibrium for a competitive
  economy, {\it Econometrica\/} {\bf 22}, 265 (1954).

\bibitem{SamuelsonNordhaus2009-2}
P.~Samuelson, W.~Nordhaus, Markets and government in a modern economy, in P.
  Samuelson, W. Nordhaus, {\it Economics\/}  (McGraw-Hill, Boston, 2008).

\bibitem{Bator1958}
F.~M. Bator, The anatomy of market failure, {\it The Quarterly Journal of
  Economics\/} {\bf 72}, 351 (1958).

\bibitem{GreenwaldStiglitz1986}
B.~Greenwald, J.~Stiglitz, Externalities in economies with imperfect
  information and incomplete markets, {\it Quarterly Journal of Economics\/}
  {\bf 90}, 229 (1986).

\bibitem{Stiglitz2009}
J.~Stiglitz, Government failure vs. market failure: Principles of regulation,
  in E. Balleisen, D. Moss (eds.), {\it Governments and markets: Toward a new
  theory of regulation\/}  (Cambridge University Press, Cambridge,
  Massachusetts, 2012).

\bibitem{Marshall1920}
A.~Marshall, The theory of monopolies, in A. Marshall, {\it Principles of
  economics\/}  (Prometheus Books, Amherst, New York, 1997).

\bibitem{Williamson1972}
O.~Williamson, Dominant firms and the monopoly problem: Market failure
  considerations, {\it Harvard Law Review\/} {\bf 85}, 1512 (1972).

\bibitem{Lande2007}
R.~H. Lande, Market power without a large market share: The role of imperfect
  information and other ``consumer protection" market failures, {\it The
  American Antitrust Institute Working Paper No 07-06\/}  (2007).
  \url{http://papers.ssrn.com/sol3/papers.cfm?abstract_id=1103613}.

\bibitem{Akerlof1970}
G.~A. Akerlof, The market for ``lemons": Quality, uncertainty, and the market
  mechanism, {\it Quarterly Journal of Economics\/} {\bf 84}, 488 (1970).

\bibitem{Cseres2006}
K.~Cseres, The impact of consumer protection on competition and competition
  law: The case of deregulated markets, {\it Amsterdam Center for Law \&
  Economics Working Paper No. 2006-05\/}  (2006).
  \url{http://papers.ssrn.com/sol3/papers.cfm?abstract_id=903284}.

\bibitem{BrixMcKee2009}
L.~Brix, K.~McKee, Consumer protection regulation in low-access environments:
  Opportunities to promote responsible finance, {\it CGAP Focus Note 60\/}
  (Washington, D.C., 2009).
  \url{http://www.cgap.org/publications/consumer-protection-regulation-low-access-environments}.

\bibitem{Arrow1969}
K.~Arrow, The organization of economic activity: Issues pertinent to the choice
  of market versus non-market allocation, in The Analysis and Evaluation of
  Public Expenditures: The PBB-System, Vol. 1, {\it U.S. Congress, Joint
  Economic Committee\/}  (Government Printing Office, Washington, D.C., 1969).
  \url{http://msuweb.montclair.edu/~lebelp/psc643intpolecon/arrownonmktactivity1969.pdf}.

\bibitem{Medema2007}
S.~Medema, The hesitant hand: Mill, {S}idgwick, and the evolution of the theory
  of market failure, {\it History of Political Economy\/} {\bf 39}, 331 (2007).

\bibitem{FullertonStavins1998}
D.~Fullerton, R.~Stavins, How do economists really think about the environment,
  {\it ENRP Discussion Paper E-98-04\/}  (Kennedy School of Government, Harvard
  University, 1998). \url{http://www.rff.org/documents/RFF-DP-98-29.pdf}.

\bibitem{Stigler1968}
G.~J. Stigler, {\it The organization of industry\/} (University of Chicago
  Press, Chicago, 1968).

\bibitem{EPA2010}
L.~P. Jackson, Remarks on the 40th anniversary of the Clean Air Act, as
  prepared, {\it Environmental Protection Agency\/}  (2010).
  \url{http://yosemite.epa.gov/opa/admpress.nsf/12a744ff56dbff8585257590004750b6/7769a6b1f0a5bc9a8525779e005ade13!OpenDocument}.

\bibitem{EPA1997}
C.~M. Browner, 25th Anniversary of the Clean Water Act, {\it Environmental
  Protection Agency\/}  (1997).
  \url{http://yosemite.epa.gov/opa/admpress.nsf/a162fa4bfc0fd2ef8525701a004f20d7/872d86a1679743df8525701a0052e3a5!OpenDocument&Highlight=0}.

\bibitem{EPA2012}
N.~Stoner, Celebrate the 40th anniversary of the Clean Water Act, {\it
  Environmental Protection Agency\/}  (2012).
  \url{http://blog.epa.gov/blog/2012/10/cwa40/}.

\bibitem{LaffontTirole1992}
J.~J. Laffont, J.~Tirole, Regulatory capture, in J. J. Laffont, J. Tirole, {\it
  A theory of incentives in procurement and regulation\/}  (MIT Press,
  Cambridge, Massachusetts, 1993).

\bibitem{McCarty2011}
N.~McCarty, Complexity, capacity, and capture, in D. Carpenter, D. Moss (eds.),
  {\it Preventing capture: Special interest influence in regulation, and how to
  limit it\/}  (Cambridge University Press, New York, 2013).

\bibitem{Woodward1998}
S.~E. Woodward, Regulatory capture at the U.S. Securities and Exchange
  Commission, {\it Milken Institute Conference on Capital Markets\/}  (Santa
  Monica, California, 1998).
  \url{http://www.sandhillecon.com/pdf/RegulatoryCapture.pdf}.

\bibitem{Krueger1974}
A.~Krueger, The political economy of the rent-seeking society, {\it American
  Economic Association\/} {\bf 64}, 291 (1974).

\bibitem{Peltzman1976}
S.~Peltzman, Toward a more general theory of regulation, {\it Journal of Law
  and Economics\/} {\bf 19}, 211 (1976).

\bibitem{Tirole1986}
J.~Tirole, Hierarchies and bureaucracies: On the role of collusion in
  organizations, {\it Journal of Law, Economics and Organization\/} {\bf 2},
  181 (1986).

\bibitem{GrossmanHelpman1993}
G.~Grossman, E.~Helpman, Protection for sale, {\it American Economic Review\/}
  {\bf 84}, 833 (1994).

\bibitem{Dixit1997}
A.~K. Dixit, G.~Grossman, E.~Helpman, Common agency and coordination: General
  theory and application to government policy making, {\it Journal of Political
  Economy\/} {\bf 105}, 752 (1997).

\bibitem{Snyder1991}
J.~M. Snyder, On buying legislatures, {\it Economics and Politics\/} {\bf 3},
  93 (1991).

\bibitem{Bo2000}
E.~D. B\'o, Bribing voters, {\it Economics Series Working Papers 9939\/}
  (University of Oxford, Department of Economics, 2000).
  \url{http://faculty.haas.berkeley.edu/dalbo/Bribing_Voters_Published_version.pdf}.

\bibitem{Bennedsen2002}
M.~Bennedsen, S.~E. Feldmann, Lobbying legislatures, {\it Journal of Political
  Economy\/} {\bf 110}, 919 (2002).

\bibitem{Neeman1999}
Z.~Neeman, The freedom to contract and the free-rider problem, {\it Journal of
  Law, Economics and Organization\/} {\bf 15}, 685 (1999).

\bibitem{Che1995}
Y.~Che, Revolving doors and the optimal tolerance for agency collusion, {\it
  RAND Journal of Economics\/} {\bf 26}, 378 (1995).

\bibitem{Overby2011}
P.~Overby, Outgoing FCC Commissioner to lobby for Comcast, {\it NPR's All
  Things Considered\/}  (2011).
  \url{http://www.npr.org/2011/05/12/136250400/for-government-employees-revolving-door-continues}.

\bibitem{Vidal2011}
J.~B. i~Vidal, M.~Draca, C.~Fons-Rosen, Revolving door lobbyists, {\it 5th
  Annual Conference on Empirical Legal Studies Paper\/}  (2011).
  \url{http://personal.lse.ac.uk/blanesiv/revolving.pdf}.

\bibitem{Axelrod1981}
R.~Axelrod, W.~D. Hamilton, The evolution of cooperation, {\it Science\/} {\bf
  211}, 1390 (1981).

\bibitem{Nowak1992}
M.~A. Nowak, K.~Sigmund, Tit for tat in heterogeneous populations, {\it
  Nature\/} {\bf 355}, 250 (1992).

\bibitem{Nowak1993}
M.~A. Nowak, K.~Sigmund, A strategy of win-stay, lose-shift that outperforms
  tit for tat in the prisoners' dilemma game, {\it Nature\/} {\bf 364}, 56
  (1993).

\bibitem{Nowak1995}
M.~A. Nowak, R.~M. May, K.~Sigmund, The arithmetics of mutual help, {\it
  Scientific American\/} {\bf 272}, 76 (1995).
  \url{http://www.ped.fas.harvard.edu/people/faculty/publications_nowak/SciAm95a.pdf}.

\bibitem{Gouldner1960}
A.~W. Gouldner, The norm of reciprocity: A preliminary statement, {\it American
  Sociological Review\/} {\bf 25}, 161 (1960).

\bibitem{Nowak1998}
M.~A. Nowak, K.~Sigmund, Evolution of indirect reciprocity by image scoring,
  {\it Nature\/} {\bf 393}, 573 (1998).

\bibitem{Berg1995}
J.~Berg, J.~Dickhaut, K.~McCabe, Trust, reciprocity, and social history, {\it
  Games and Economic Behavior\/} {\bf 10}, 122 (1995).

\bibitem{Fehr2000}
E.~Fehr, S.~Gachter, Fairness and retaliation: The economics of reciprocity,
  {\it The Journal of Economic Perspectives\/} {\bf 14}, 159 (2000).

\bibitem{Clark2001}
K.~Clark, M.~Sefton, The sequential prisoners' dilemma: Evidence on
  reciprocation, {\it The Economic Journal\/} {\bf 111}, 51 (2001).

\bibitem{Lambsdorff2010}
J.~G. Lambsdorff, Deterrence and constrained enforcement -- Alternative regimes
  to deal with bribery, {\it Passauer Diskussionspapiere\/} {\bf 60}
  (UniversitŠt Passau, 2010).
  \url{http://www.wiwi.uni-passau.de/fileadmin/dokumente/lehrstuehle/wilhelm/Working_Papers_PDF/Disk_60_Deterrence_and_Constrained_Enforcement.pdf}.

\bibitem{Schelling1960}
T.~C. Schelling, {\it The strategy of conflict\/} (Oxford University Press,
  Oxford, 1960).

\bibitem{Carpenter2011}
D.~Carpenter, Detecting and measuring capture, in D. Carpenter, D. Moss (eds.),
  {\it Preventing capture: Special interest influence in regulation, and how to
  limit it\/}  (Cambridge University Press, forthcoming 2013).

\bibitem{Kahan1999}
D.~M. Kahan, E.~A. Posner, Shaming white-collar criminals: A proposal for
  reform of the federal sentencing guidelines, {\it Journal of Law and
  Economics\/} {\bf 42}, 365 (1999).

\bibitem{Barr2001}
A.~Barr, Social dilemmas and shame-based sanctions: Experimental results from
  rural Zimbabwe, {\it Economics Series Working Papers WPS/2001-11, University
  of Oxford, Department of Economics\/}  (2001).
  \url{http://ideas.repec.org/p/csa/wpaper/2001-11.html}.

\bibitem{Schulze2003}
G.~G. Schulze, B.~Frank, Deterrence versus intrinsic motivation: Experimental
  evidence on the determinants of corruptibility, {\it Economics of
  Governance\/} {\bf 4}, 143 (2003).

\bibitem{Panagopoulos2010}
C.~Panagopoulos, Affect, social pressure and prosocial motivation: Field
  experimental evidence of the mobilizing effects of pride, shame and
  publicizing voting behavior, {\it Political Behavior\/} {\bf 32}, 369 (2010).

\bibitem{Krain2012}
M.~Krain, J'accuse! Does naming and shaming perpetrators reduce the severity of
  genocides or politicides?, {\it International Studies Quarterly\/} {\bf 56},
  574 (2012).

\bibitem{Bar-Yam2005}
Y.~Bar-Yam, {\it Making things work\/} (NECSI Knowledge Press, Cambridge,
  Massachusetts, 2005).

\bibitem{Lagi2011a}
M. Lagi, Yavni Bar-Yam, K.Z. Bertrand, Yaneer Bar-Yam, The food crises: A
  quantitative model of food prices including speculators and ethanol
  conversion, {\it arXiv:1109.4859\/}  (2011).
  \url{http://necsi.edu/research/social/foodprices.html}.

\bibitem{Lagi2011b}
M.~Lagi, K.~Bertrand, Y.~Bar-Yam, The food crises and political instability in
  North Africa and the Middle East, {\it arXiv:1108.2455\/}  (2011).
  \url{http://necsi.edu/research/social/foodcrises.html}.

\bibitem{House2004}
Clerk of the House, {\it H.R. 4520 (108th) American Jobs Creation Act: House
  vote no. 509\/}  (Washington, D.C., 2004).
  \url{http://www.govtrack.us/congress/votes/108-2004/h509}.

\bibitem{Senate2004}
Senate Bill Clerk, {\it H.R. 4520 (108th): Senate vote no. 211\/}  (Washington,
  D.C., 2004).
  \url{http://www.senate.gov/legislative/LIS/roll_call_lists/roll_call_vote_cfm.cfm?congress=108&session=2&vote=00211}.

\bibitem{housepay}
I.~A. Brudnick, Salaries of members of Congress: Recent actions and historical
  tables, {\it Congressional Research Service\/}  (Washington, D.C., 2013).
  \url{http://www.senate.gov/CRSReports/crs-publish.cfm?pid='*2%404P%5C%5B%3A%22%40%20%20%0A}.

\bibitem{senatepay}
I.~A. Brudnick, Congressional salaries and allowances, {\it Congressional
  Research Service\/}  (Washington, D.C., 2012).
  \url{http://library.clerk.house.gov/reference-files/112_20120104_Salary.pdf}.

\bibitem{Rose-Ackerman1978}
S.~Rose-Ackerman, {\it Corruption -- A study in political economy\/} (Academic
  Press, New York, 1978).

\bibitem{Rose-Ackerman1999}
S.~Rose-Ackerman, {\it Corruption and government: Causes, consequences, and
  reform\/} (Cambridge University Press, New York, 1999).

\bibitem{Shleifer1993}
A.~Shleifer, R.~W. Vishny, Corruption, {\it Quarterly Journal of Economics\/}
  {\bf 108}, 599 (1993).

\bibitem{Glaesera2006}
E.~L. Glaesera, R.~E. Saks, Corruption in America, {\it Journal of Public
  Economics\/} {\bf 90}, 1053 (2006).

\bibitem{Epstein1999}
D.~Epstein, S.~O'Halloran, {\it Delegating powers: A transaction cost politics
  approach to policy making under separate powers\/} (Cambridge University
  Press, New York, 1999).

\bibitem{Puzzanghera2008}
J.~Puzzanghera, W.~Hamilton, Critics assail watchdog's lax oversight, {\it Los
  Angelos Times\/}  (2008).
  \url{http://articles.latimes.com/2008/dec/18/business/fi-madoff-sec18}.

\bibitem{Lawson2011}
G.~Lawson, The stoner arms dealers: How two American kids became big-time
  weapons traders and how the Pentagon later turned on them, {\it Rolling
  Stone\/}  (2011).
  \url{http://www.rollingstone.com/politics/news/the-stoner-arms-dealers-20110316}.

\bibitem{Taibbi2011a}
M.~Taibbi, The People vs. Goldman Sachs, {\it Rolling Stone\/}  (2011).
  \url{http://www.rollingstone.com/politics/news/the-people-vs-goldman-sachs-20110511}.

\bibitem{Taibbi2011b}
M.~Taibbi, Is the SEC covering up Wall Street crimes?, {\it Rolling Stone\/}
  (2011).
  \url{http://www.rollingstone.com/politics/news/is-the-sec-covering-up-wall-street-crimes-20110817}.

\bibitem{Ferguson2012}
C.~Ferguson, Heist of the century: Wall Street's role in the financial crisis,
  {\it The Guardian\/}  (2012).
  \url{http://www.theguardian.com/business/2012/may/20/wall-street-role-financial-crisis}.

\end{thebibliography}

\end{document}